# Noise Pulsing of Narrow-Band ASE from Erbium-Doped Fiber Amplifier

Pablo Muniz-Cánovas, Yuri O. Barmenkov, *Member, IEEE*, Alexander V. Kir'yanov, *Member, IEEE*, José L. Cruz, and Miguel V. Andrés, *Member, IEEE*

*Abstract*—In this paper, we report an experimental study of noise features of polarized and unpolarized amplified spontaneous emission (ASE) with narrow optical bandwidth, registered from a conventional low-doped erbium fiber. We demonstrate that ASE noise can be considered as train of Gaussian-like pulses with random magnitudes, widths, and inter-pulse intervals. The statistical properties of these three parameters of noise pulsing are analyzed. We also present the results on the influence of ASE noise upon optical spectrum broadening, produced by self-phase modulation at propagating along communication fiber, and demonstrate that ASE noise derivation stands behind the broadening shaping.

## I. INTRODUCTION

Amplified spontaneous emission (ASE) based light sources are characterized by broad optical spectrum and high temporal stability given the absence of relaxation oscillations and interference effects. Such kind of light sources are successful in many applications, including high-precision fiber-optic gyroscopes [1,2], low-coherence interferometry [3], optical coherence tomography [4,5], *etc*. Recently, it was demonstrated that nanosecond ASE pulses produced by actively Q-switched fiber lasers may serve as effective pump for supercontinuum generation [6-8]. ASE sources may also serve as seed for CW amplifying to a multi-hundred watts level with simultaneous suppressing of SBS issues because of the broad ASE spectrum [9-11]. Note that, in fiber amplifiers used in fiber-optic links, ASE noise is naturally added to amplified signal, which deteriorates signal-to-noise ratio [12-15].

ASE photon noise is described by *M*-fold degenerate Bose-Einstein distribution, where *M* corresponds to the number of independent states (modes) of ASE [16,17], defined by ratio of ASE optical spectrum width ($B_{opt}$) to photodetector electric bandwidth ($B_{el}$) and polarization degeneracy (*s*). The basic properties of ASE noise statistics are known; nevertheless, its fine features, characteristic to ASE random photon pulsing (shown, for instance, in Fig. 1, Ref. [18]), present certain interest.

In this paper, we report the experimental data on random ASE noise pulsing inherent to an erbium-doped fiber amplifier (EDFA) at optical filtering by fiber Bragg gratings (FBGs) of different spectral widths. The set of FBGs used in the experiments and the available photo-receiving equipment permitted us to vary the ratio $B_{opt}/B_{el}$ from 0.16 to 9.3; the number of orthogonal polarization states *s* was set to one or two, per demand.

We demonstrate that noise pulses of which ASE signal is composed are Gaussian-like, with magnitudes, widths, and sequencing intervals described by specially parameterized distributions. We also show that the distribution of ASE signal's time derivative, causing optical spectrum broadening through self-phase modulation in optical fiber, has symmetric triangular-like shape at semi-logarithmic scaling. The data on spectral broadening of ASE signal passed through a long communication fiber confirms this statement.

## II. EXPERIMENTAL SETUP

Our experimental setup is shown in Fig. 1. It comprises a seed ASE source based on erbium-doped fiber (EDF) EDF1 and a fiber amplifier based on EDF2. The EDF used was a standard low-doped M5-980-125 C-band fiber with small-signal gain of ~ 6.5 dB/m at 1530 nm. Both EDFs were pumped by commercial diode lasers at 976 nm through fused 976/1550 nm wavelength division multiplexers (WDMs). EDF1 and EDF2 lengths were approximately 6 m each; this length provided relatively high ASE power and, at the same time, prevents parasitic CW lasing that otherwise may arise at 1530 nm due to very weak reflections from fiber circulators, WDMs, and 7°- fiber cuts when the active fiber is long [19]. For the same purpose, a long-period grating (LPG) with attenuation peak at 1530 nm was utilized in the scheme. Circulators (C1 and C2) served for preventing feedbacks between the seed ASE source, the fiber amplifier, and the output fiber patch cable with PC/APC termination. A fiber polarizer (POL) was placed at the output of the ASE source for studying the properties of polarized ASE; otherwise it was

Manuscript received August 1, 2018. This work was partially supported by the Agencia Estatal de Investigación (AEI) of Spain and Fondo Europeo de Desarrollo Regional (FEDER) (Ref. TEC2016–76664-C2-1-R).
Pablo Muniz-Cánovas, Y. O. Barmenkov, and A. V. Kir'yanov are with the Centro de Investigaciones en Optica, Leon 37150, Mexico (e-mails: pablomc@cio.mx; yuri@cio.mx; kiryanov@cio.mx); A. V. Kir'yanov is also with the National University of Science and Technology "MISIS", Moscow 119049, Russian Federation.
J. L. Cruz and M. V. Andres are with the Departamento de Fisica Aplicada, Instituto de Ciencia de Materiales, Universidad de Valencia, 46100 Valencia, Spain (e-mails: jose.l.cruz@uv.es; miguel.andres@uv.es).

removed.

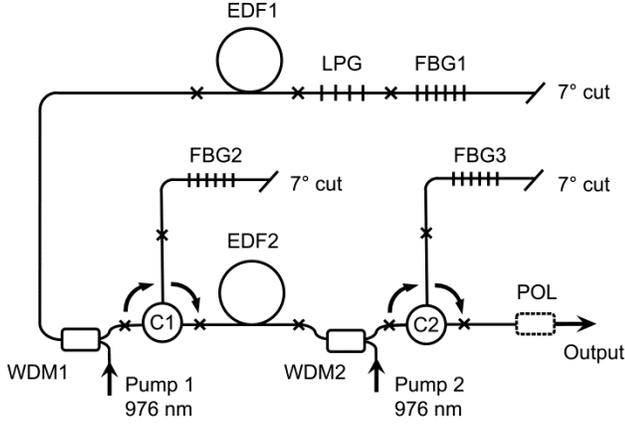

Fig. 1. Experimental arrangement of ASE source; crosses indicate fiber splices.

To select an operation wavelength of the ASE source and to vary its optical bandwidth, a set of home-made FBGs centered at 1544.6 nm was fabricated. The gratings FBG1 and FBG2 were broadband (~ 600 pm) while the grating FBG3 (replaceable) was of a narrower band: in fact, it defined the spectrum width of ASE signal under statistical study. A collection of the ASE spectra, obtained using the gratings FBG3 with different bandwidths, is presented in Fig. 2; in turn, ASE bandwidths measured at a 3-dB level (FWHM) are shown in the Table on its right side.

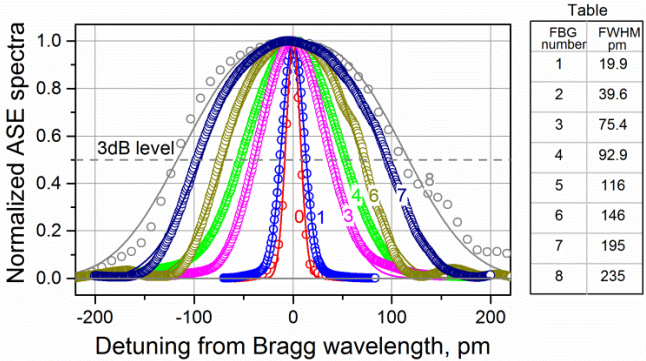

Fig. 2. Normalized ASE spectra measured for different FBGs (FBG3). Table indicates FWHMs of ASE spectra. Curve "0" demonstrates OSA response to narrow-line laser signal (optical width, 130 kHz). Circles – experimental points; lines – either Gaussian fits (curves 0 to 4) or fits by sub-Gaussian function with power varied in the range 2.7 to 2.9 (curves 5 to 8). Some of FBG spectra are not shown to keep clarity of the picture.

To measure ASE optical spectra, we employed an optical spectrum analyzer (OSA) with 17.2 pm-resolution within the erbium window (*Yokogawa*, model AQ6370B); OSA resolution was estimated from its Gaussian-like response to a narrow-line (130 kHz) laser signal: see the curve marked as "0" in Fig. 2. This resolution was used for correcting widths of the narrow-band Gaussian-like ASE spectra (curves 1 to 4) by deconvolution of the two spectra (of OSA and of a FBG3 used).

To record ASE signals, two sets of devices were implemented: the first set, included 5-GHz InGaAs p-i-n photodetector (*Thorlabs*, model DET08CFC) and 3.5-GHz-oscilloscope (*Tektronix*, model DPO7354C) with RF band extending from DC to factual 3.3GHz, and the second one, based on 25-GHz InGaAs Schottky photodetector (*Newport*, model 1414) and 16-GHz real-time oscilloscope (*Tektronix*, model DPO71604C) with overall RF band extending from DC to factual 15.5 GHz; in both realizations, RF band was measured at a 3-dB level. In all experiments, output ASE power was set to ~ 2 mW, a value that insured functioning of the photodetectors well below saturation.

### III. EXPERIMENTAL RESULTS AND DISCUSSION

First, we checked whether the statistics of ASE signals are properly described by the *M*-fold degenerate Bose-Einstein distribution. We measured histograms for all available optical widths of both polarized and unpolarized ASE and for each of the two photo-registration schemes, described above. Second, all experimental histograms were fitted by curves, simulated using the formula for Bose-Einstein distribution [16,17]:

$$P(n, \bar{n}, M) = \frac{(n+M-1)!}{n!(M-1)!} \frac{(\bar{n})^n}{(1+\bar{n})^{n+M}} \qquad (1)$$

where $P(n, \bar{n}, M)$ is the probability of counting *n* photons by photodetector during counted (average) time $T=1/B_{el}$ and $\bar{n}$ is the mean photon count in the same time interval. Here *M* is number of independent ASE states (modes), which, for the Gaussian optical spectrum, is defined as [16,17]:

$$M = s \frac{\pi (B_{opt}/B_{el})^2}{\pi (B_{opt}/B_{el}) \text{erf}[\sqrt{\pi}(B_{opt}/B_{el})] - [1-\exp(-\pi(B_{opt}/B_{el})^2)]} \qquad (2)$$

Note that, for $M = 1$ and large $\bar{n}$, formula (1) is approximated by an exponential decay function:

$$P(n, \bar{n}, M = 1) \approx \frac{1}{\bar{n}} \exp\left(-\frac{n}{\bar{n}}\right) \propto \frac{1}{I_0} \exp\left(-\frac{I}{I_0}\right) \qquad (3)$$

where *I* and $I_0$ are the light intensity and its mean value, respectively. Note that ASE, approximated by the exponential probability density function (PDF) in accord with (3), is Gaussian for the field, which is assumed, for instance, in wave turbulence theory for the weakly nonlinear regime [20,21].

As seen from Fig. 2, only the four narrowest ASE spectra are nearly Gaussian (curves 1 to 4), whereas the other ones are broader and more flattened (of sub-Gaussian appearance) with *M*-values intermediate between the ones found for the Gaussian and the rectangular spectra (see Fig. 6-1 in [16]). Note that the difference between these two limiting values dramatically decreases with increasing optical spectrum's width (and so with increasing *M*).

In Fig. 3, we present six examples of experimental histograms of ASE noise, obtained for the ratios $B_{opt}/B_{el}$ equal to 0.16 (Figs. (a1) and (a2)), to 0.75 (Figs. (b1) and (b2)), and to 2.9 (Figs. (c1) and (c2)), and their fits. The left column (1) of the figure corresponds to linearly polarized light ($s = 1$) while the right column (2) to unpolarized light ($s = 2$). In all six panels, the horizontal axis is normalized to the mean photon number (or the mean voltage for the experimental points) and the experimental noise count (proportional to photon count) is recalculated to probability. For each case, the fitting curves obtained using Eq. (1) and the best fits for $M$-value are demonstrated by the green lines. Besides, for $B_{opt}/B_{el} < 1$, we add the histograms simulated for the 'ideal' cases when $M = 1$ (polarized light, the left column) and $M = 2$ (unpolarized light, the right column), shown by the red lines. The fitting curves were simulated using the mean photon number $\bar{n} = 5.3 \times 10^6$ for the 3.3-GHz detection scheme and $\bar{n} = 1.1 \times 10^6$ for the 15.5-GHz one.

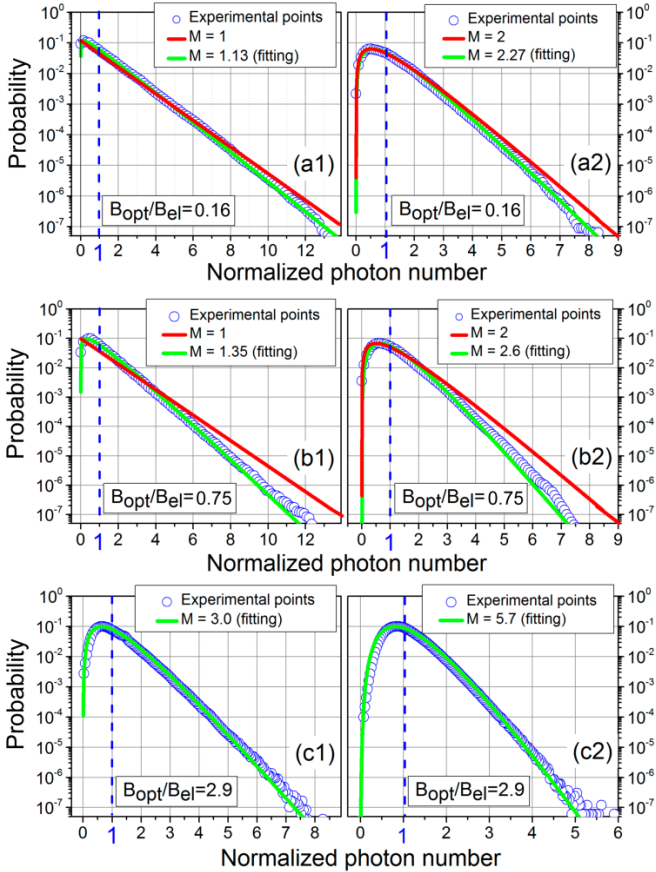

Fig. 3. Histograms of probability (PDF) of the normalized photon number (ASE noise voltage) for three different values of $B_{opt}/B_{el}$ ratio. Histograms obtained for polarized ($s=1$) and unpolarized ($s=2$) ASE are shown in the left and right column, respectively. Circles are experimental points, green lines are the best fits, and red lines are simulated probabilities for the ideal cases: $M = 1$ at $s=1$ and $M = 2$ at $s = 2$.

As seen from Fig. 3, the experimental histograms are properly described by $M$-fold degenerate Bose-Einstein distributions at $M$ providing the best fits; this reveals that ASE light in our case is a classical thermal source. As is also seen, for the ratios $B_{opt}/B_{el}$ less or even much less than unity (Fig. 3, (a1) to (b2)) the mode number $M$ is always bigger than 1 and 2 for, correspondingly, polarized and unpolarized ASE, given that the absolute slope values for the red curves are always less than those for the fitting lines.

Fig. 4 shows the experimental $M$-values in function of $B_{opt}/B_{el}$ (symbols) and the corresponding fitting curves, simulated by Eq. (2) for polarized ($s=1$, curve 1) and unpolarized ($s=2$, curve 2) ASE; as seen, the modeling data perfectly fit the experimental ones. Note that $M$-values obtained for unpolarized ASE are two times larger than those obtained for polarized ASE, which confirms that the mode numbers differ by two times in these two cases.

Furthermore, it is seen from Figs. 3 and 4 that, if the ratio $B_{opt}/B_{el}$ is less than 1, the mode number $M$ is always >1 and >2, for polarized and unpolarized ASE, respectively. Note here that in the ideal circumstances ($M = 1$ or $M = 2$ for polarized/unpolarized light), optical bandwidth is null ($B_{opt} = 0$), which corresponds to practically inaccessible "single-frequency" ASE. On the contrary, if $B_{opt}/B_{el}$ is very high, the mode number equals to this ratio multiplied by $s$.

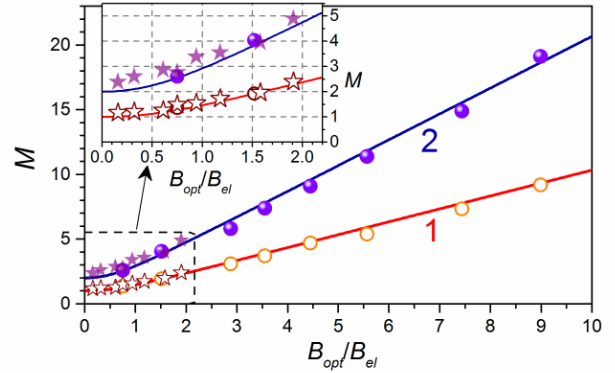

Fig. 4. Mode number $M$ vs. ratio $B_{opt}/B_{el}$ ratio. Symbols are experimental points; lines are simulated using Eq. 2. Lines 1 and 2 (with symbols nearby) correspond to polarized and unpolarized ASE, respectively. Circles and stars are obtained using 3.3-GHz and 15.5-GHz detection setups, respectively. The initial part of the dependencies (see the dashed rectangle in the main window) is zoomed in the inset.

Below we limit ourselves by addressing the ASE features arising at lowest $B_{opt}/B_{el}$, at which the histograms of photon count are broadest (*viz.* narrowest optical spectrum and broadest RF spectrum), since an increase of this ratio may lead to strong frequency distortion of ASE signal through dramatic dumping of the high-frequency ASE components.

Fig. 5 presents two examples of randomly chosen short sections of the oscilloscope traces selected from the long ones, captured for polarized ASE noise with (a) 3.3-GHz and (b) 15.5-GHz detection schemes. These traces were obtained under the same conditions at which the histograms shown in Figs. 3(b1) and 3(a1) were recorded (i.e. $M = 1.35$ and $M = 1.13$, respectively). In the figure, circles are the experimental points, triangles are the peaks' magnitudes, and the solid lines are the Gaussian fits of noise peaks. One can notice from Fig. 5 that ASE noise may be considered as train of noise pulses with random magnitudes, widths, and intervals of sequencing and that each pulse may be fitted by Gaussian function, which permits determining its width and position in train.

Fig. 6 shows the histograms of probability of magnitudes of ASE noise pulses obtained from the long oscilloscope traces (31.25 MS that was a limit for 16 GHz-oscilloscope), for both polarized and unpolarized ASE with the narrowest $B_{opt}$ and broadest RF-bandwidth (15.5 GHz). In the figure, the stars stand for the experimental probabilities of the noise peak magnitudes, normalized to the mean value. It is seen that these two dependences match well the ones simulated for the ideal Bose-Einstein distributions with $M = 1$ and $M = 2$ (for polarized and unpolarized light, respectively) with a sole difference that they are slightly shifted up with respect to the ideal circumstances, demonstrated by the dash orange lines.

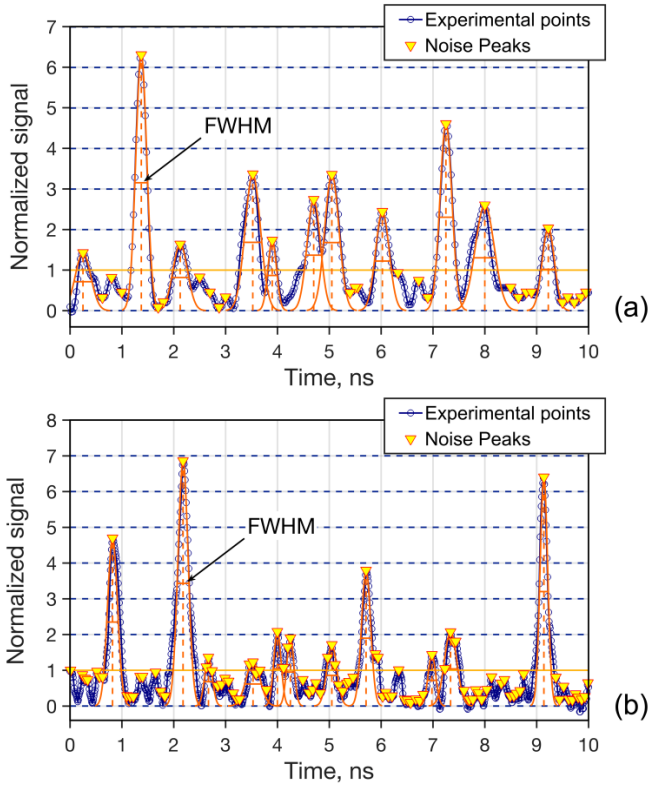

Fig. 5. ASE noise as train of Gaussian-like pulses (circles); ASE signals are normalized to the mean values of detector signals. Vertical dash lines indicate centers of Gaussian fits; here, the pulses with magnitude less than the mean are not fitted. The intervals between adjacent points are (a) 25 ps and (b) 10 ps.

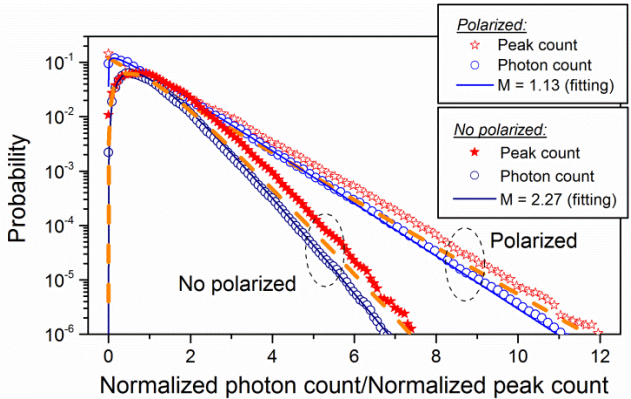

Fig. 6. Histograms measured for magnitudes of ASE peaks (stars) and photon counts (circles), for $B_{opt}/B_{el} = 0.16$. Both photon counts and ASE peaks are normalized to the mean photon count.

For comparison, probabilities of photon counts, together with the best fits, are presented in Fig. 6, too. As seen, the histograms of the peak counts are broader than those for the photon counts, the feature observed for both polarized and unpolarized ASE. This effect seems to stem from the fact that in the former case all experimental points below ASE peaks were not considered. Note that occasional excessive noise peaks with magnitudes much larger than the mean power may be treated as rogue waves or their seeds [22-24].

Another important point, regarding train of ASE pulses, is dispersion of the intervals between pulses belonging to different ranges of magnitudes (*P*) (refer to Fig. 5). Such ranges were identified in the following manner: (*i*) below the mean photon count (*m*), (*ii*) from the mean to two means, (*iii*) from two to three means, *etc.*, but with a limit $P = 9m$ to $10m$ (at a higher peak magnitude, the peak counts are too small for proceeding a statistical study).

Fig. 7 shows two examples of histograms built for intervals between ASE pulses for two different ranges of magnitudes (specified in panels (a) and (b)) and single polarization state. The histograms are fitted with a high confidence by a linear dependence in the semi-logarithmic scale, revealing exponential decay of the peak count *vs.* interval between pulses.

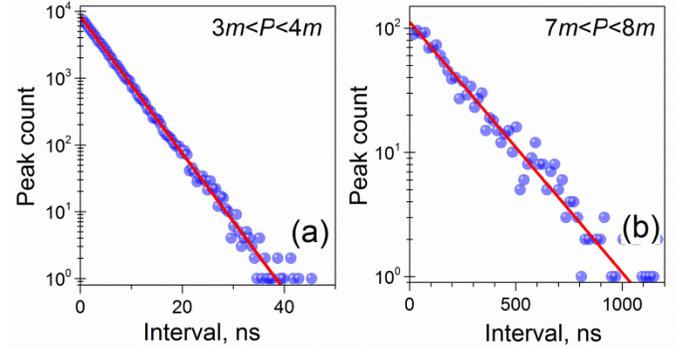

Fig. 7. Examples of histograms for intervals between ASE pulses, belonging to the ranges of the normalized peak magnitudes limited to (a) $3m$ to $4m$ and (b) $7m$ to $8m$; $B_{opt}/B_{el} = 0.16$; ASE is polarized: $M = 1.13$ (symbols, experimental data; lines, exponential fits).

Fig. 8(a) resumes the fits of the peak count "decay", found for all considered cases in log-log scale, whilst Fig. 8(b) shows 3D dependence of peak count normalized to its maximum (in color scaling) in function of the normalized peak magnitude and inter-pulse interval (in the figure – "peaks' separation") of the same range of magnitudes *P*.

It is seen from Fig. 8 that probability of shorter intervals between adjacent peaks of the same range of magnitudes is higher, while that of longer ones fades exponentially with peaks' enlarging. The higher magnitude of ASE peaks, the decay is slower. Note that in this figure the minimal peak separation is fixed to 175 ps, corresponding to the width of a transform-limited Gaussian pulse with optical spectrum measured by 20 pm (2.5 GHz in frequency domain) and defining the minimal separation of two Gaussian peaks to be resolved (see, for example, Ref. [25]).

Then, note that the peak count at minimal peaks' separation decreases by exponential law with increasing the peak magnitude (see the vertical axis in Fig. 8(a)) whereas the "cut-off value of the peaks' separation, measured at -3dB below the maximum peak count, increases, also by exponential law, with increasing *P*: see the diagonal straight line in Fig. 8(b). This line splits the graph surface in two triangular areas with maximal (red color) and minimal (blue color) values of the normalized peak count, separated by narrow transitional area (of approximately an order of magnitude of peaks' separation).

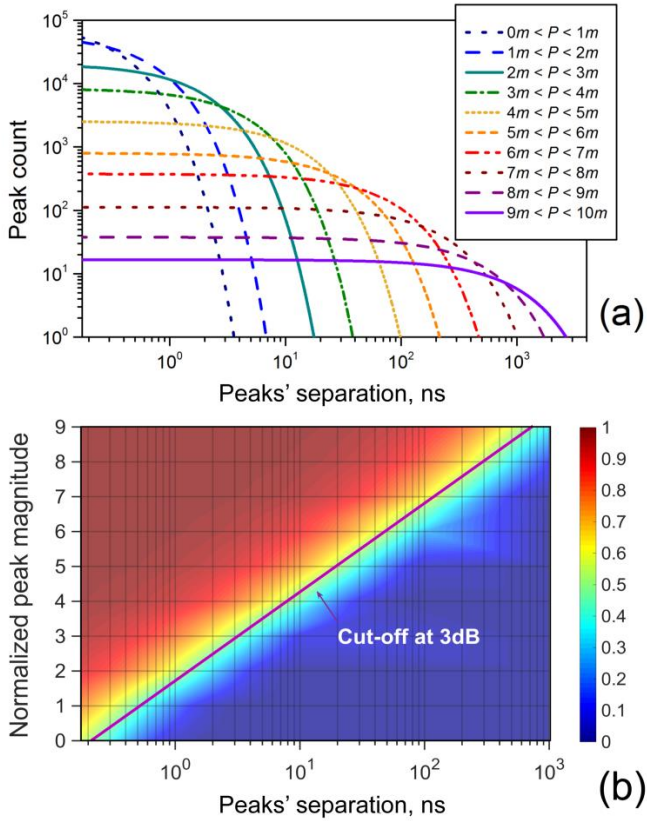

Fig. 8. (a) Peak count (absolute values) and (b) peak count normalized to its maximum (color scale) in function of normalized peak magnitude and interval between peaks of the same ranges of magnitude.

In Fig. 9, are shown the histograms of width at FWHM ($\Delta t_{FWHM}$) of polarized ASE, parameterized for different $P$-ranges. As seen, both the shapes of the histograms and the most probabilistic widths depend on pulse magnitude. Interestingly, with increasing $P$, the most probabilistic width increases, too. It is also worth noticing that shapes of all the histograms obtained for $P < 5m$ are very similar, only differing in the most probable width (compare the two upper curves); in the meantime, for higher $P$, the histograms become more symmetrical (see the lower curve).

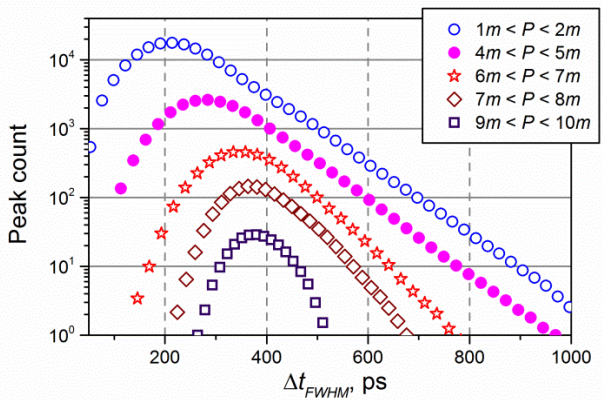

Fig. 9. Examples of histograms of ASE pulses' width, obtained for different normalized peak magnitudes $P$.

The two extreme types of the histograms are shown in more detail in Fig. 10; these are presented in both time- (left column) and frequency- (right column) domains (in the last case, the time scale was recalculated as $\Delta \nu_{FWHM} = 0.441/\Delta t_{FWHM}$, according to the Fourier transformation of a Gaussian pulse).

It is seen from Fig. 10 that, while magnitudes of the pulses are small (1$m$ to 2$m$, see the upper panels (a1) and (a2)), the right slopes of the histograms (presented in either domain) are perfectly fitted by linear dependences, adhering the exponential law of decay with increasing both $\Delta t_{FWHM}$ and $\Delta \nu_{FWHM}$. Accordingly, the exponential decay in the frequency-domain transforms to the logarithmic growth when presented in the time-domain, thus the left parts of the histograms shown are fitted by the logarithmic functions. In contrast, at a high magnitude of pulses, *viz.* an order higher than the mean count $m$, the distributions of ASE pulse widths in both domains are virtually symmetrical and fitted well by the Gaussian functions (see the lower panels (b1) and (b2)).

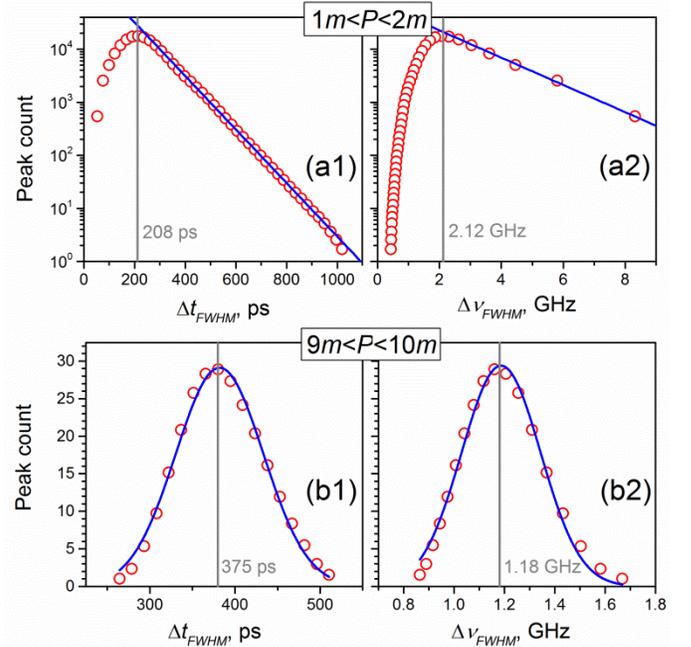

Fig. 10. Examples of histograms of ASE pulses' width, obtained for two different normalized peak magnitudes $P$.

Fig. 11 summarizes the results of statistical treatment of width of ASE noise pulses, assessed on half of magnitude. As seen from (a), with increasing the pulse magnitude from 1$m$ (1$m$<$P$<2$m$) to 9$m$ (9$m$<$P$<10$m$) the most probabilistic pulse width increases by approximately two times (see curve 1, left scale) though its absolute deviation measured on a -3dB level (see curve 2, right scale) decreases by ~40%. In turn, see (b), the relative deviation of pulse width drops stronger, by ~3 times.

Thus, the histograms for the high-magnitude ASE pulses are much narrow as compared with those for the low-magnitude pulses. We expect that the histograms of pulse widths with magnitudes exceeding the examined ones ($P$>10$m$) should be yet more symmetrical and narrow and, hence, more uniform in shape and less deviated from the most probabilistic value.

Let now snapshot an ASE noise feature of nonlinear nature. An interesting property of ASE noise concerns the nonlinear optical phenomena. It is known that, given that the broadening magnitude is proportional to the optical power derivative [26], a derivative of an optical signal causes the effect of self-phase modulation (SPM). Therefore, a derivative of ASE noise should merely affect the nonlinear effects in an optical fiber, pumped by a source of such type.

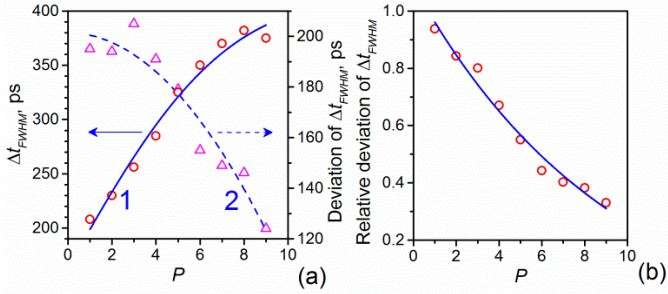

Fig. 11. (a) Width of ASE pulses (solid line, left scale) and its absolute deviation (dashed line, right scale) *vs*. normalized peak magnitude $P$; (b) relative pulse width deviation *vs.* $P$ (symbols, experimental data; lines, fits).

Fig. 12(a) demonstrates the histograms of the normalized derivative of large train of ASE pulses (its short sections are shown in Fig. 5), found as $\Delta V/\Delta t/V_{mean}$, for both polarized and unpolarized ASE, where $\Delta V$ is the voltage difference between two adjacent experimental points, $\Delta t = 10$ ps is the interval between them (refer to Fig. 5(b)), and $V_{mean}$ is the mean voltage.

One can see from this figure that the derivatives of both polarized and unpolarized ASE signals have triangular shape in semi-logarithmic scale (or exponential decays in linear scale), symmetric with respect to the vertical dash line (within at least ~60-dB range); width of the histogram for polarized ASE is by about two times greater than that for unpolarized ASE. The triangular shape of the derivative itself results from the linear distribution of magnitudes of ASE peaks, shown in Fig. 6 in semi-logarithmic scale.

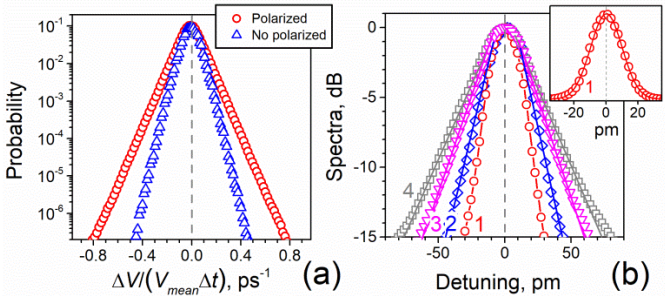

Fig. 12. (a) Histogram of normalized derivative of polarized (circles) and unpolarized (triangles) ASE noise. (b) Normalized spectrum of ASE signal on input of long communication fiber (line 1) and broadened normalized spectra measured at the fiber output (lines 2 to 4); all spectra are normalized to maxima. Symbols are experimental points and solid lines are fits. Inset shows the spectrum of input ASE signal in lineal scale (circles) and Gaussian fit (line). In panel (a) and inset, detuning is given respectively to ASE peak wavelength (~ 1544.6 nm).

Fig. 12(b) demonstrates broadening of ASE spectrum after propagating 20 km of a standard communication fiber (SMF-28); ASE polarization in this case was random. In the experiment, a scalable optical amplifier of the "seed" ASE source (discussed above) was used, which permitted us to scale ASE power up to 82 mW while keeping unchanged the ASE spectrum width. If one considers 100 ps as the lowest pulse-width limit (see Fig. 10(a1)), the nonlinear length of the fiber is ~10 km or less and the dispersion length is ~500 km; thus, the nonlinear effects appear to be much stronger than the broadening of ASE pulses because of fiber dispersion.

Curve 1 in fig. 12(b) (copied in the inset) shows the spectrum of ASE signal at the communication fiber input; its shape is clearly Gaussian with the same width as that in Fig. 2. Curves 2 to 4 show the ASE spectra at the fiber output for input powers of 30, 52, and 82 mW, respectively; it is evident steady and symmetrical broadening of the spectra with increasing ASE power.

The broadened spectra, likewise the histograms of ASE signal derivative (refer to Fig. 12(a)), are characterized by almost triangular profile in semi-logarithmic scale. Slight discrepancy from the triangularity, observed for maximal ASE power (82 mW) below -13 dB, is produced – through modulation instability – by small rudiments of the peaks (not shown). Note that, for our experimental conditions, an estimated nonlinear phase shift $\phi_{NL}$ arising due to SPM mainly matches the range extending from zero to 10 rad (at $P$ varied from $0m$ to $10m$) or covers even larger values for extremely rare events; at $P = P_{mean}$ $\phi_{NL}$ is equal to 1 rad. Also note that ASE spectral broadening in optical fiber due to SPM and its dispersion were recently modelled [17] in assumption that ASE pump is a harmonically modulated signal of some average frequency and magnitude.

Thus, by means of the last demonstration, we reveal that the spectral broadening of relatively low-power ASE signal ($P_{mean}$ < 100 mW), propagating along fiber, inheres to the triangular in semi-logarithmic scale distribution of the noise derivative.

## IV. CONCLUSION

In this paper, we presented the experimental data on noise features of amplified spontaneous emission (ASE), outcoming from standard low-doped erbium fiber. ASE was optically filtered using a set of fiber Bragg gratings (FBGs) with reflection spectra varied in a broad range, of about an order of width. The spectra obtained using FBGs with narrow spectra are nearly Gaussian while those with the broadest spectra are flatten sub-Gaussian.

First, we proved that ASE photon statistics is described by $M$-fold degenerate Bose-Einstein distribution with $M$ dependent on the optical spectrum width and number of polarization states, eventually confirming the thermal kind of such light source. We also demonstrated that, for polarized ASE with optical spectrum much narrower than RF band of photodetector, the degeneracy factor, or mode number, $M$ is always above unity and, hence, the photon statistics newer obeys an exponential law.

Second, we revealed some specific features of ASE noise for the narrowest available optical spectra (2.5 GHz) when mode number $M$ equals to 0.16 for polarized light. It was shown that ASE noise may be represented by train of Gaussian-like pulses with randomly distributed magnitudes, widths, and sequencing intervals. It was found that the probability distributions of pulse magnitudes are slightly displaced as compared with the "ideal" cases ($M = 1$ and $M =$ 2 for polarized and unpolarized light, respectively) but with slopes kept nearly the same. Besides, count of intervals between ASE pulses of the same magnitude fades by exponential law with increasing the interval between pulses: the bigger pulse magnitude, the slower decay is.

Third, we characterized in more detail the distributions of width of ASE pulses, in function of their magnitudes. For pulse magnitudes lower than four-means of ASE power, the histograms are strongly asymmetrical and broad and their left right slopes are described in time domain by the logarithmic and exponential functions, respectively (both at linear scaling). At increasing the pulse magnitudes, the histograms become narrower and more symmetric; besides, for pulses of high magnitude (bigger than nine-means of ASE power), the histograms match Gaussian distributions. Note that the most probabilistic pulse width increases with pulse magnitude (from ~200 ps for smallest pulses to ~400 ps for pulses of an order of magnitude bigger than the mean power). In other words, more powerful ASE pulses are more stable in width than less powerful ones.

Finally, we assessed the influence of ASE pulsing upon broadening of the optical spectrum at propagating along a communication fiber. We demonstrated that the shape of broadening at the fiber output is defined by the shape of ASE noise derivative (triangular at semi-logarithmic scaling), which is characteristic to SPM at weak nonlinearity.

ACKNOWLEDGMENTS

P. Muniz-Cánovas acknowledges financial support from the CONACyT (Mexico) for his Ph.D. study; A.V. Kir'yanov acknowledges financial support via the Increase Competitiveness Program of NUST «MISIS» of the Ministry of Education and Science (Russian Federation) under Grant K3-2017-015 for his sabbatical stay; J.L. Cruz and M.V. Andrés acknowledge financial support from the Agencia Estatal de Investigación (AEI) of Spain and Fondo Europeo de Desarrollo Regional (FEDER) (Ref. TEC2016–76664-C2-1-R).

**Pablo Muniz-Cánovas** received B.S. degree in Electronics and Communications Engineering from the University of Guanajuato, Guanajuato, Mexico, in 1998, and M.D. in Information Technology from Mexico Autonomous Institute of Technology, Mexico City, Mexico, in 2005. He worked for 17 years in private sector in TICs companies such as Nokia, NAVTEQ and T-Systems and holds one patent. Actually he is Ph.D. student at the Centro de Investigaciones en Optica, A.C. His current research interests concern fiber Bragg and long period gratings, fiber lasers, and fiber laser noise.

**Yuri O. Barmenkov** received his Ph.D. degree in radio-physics and electronics from the St. Petersburg State Technical University, St. Petersburg, Russia, in 1991. He was a Professor Assistant and then a Senior Lecturer with the Department of Experimental Physics of this University, from 1991 to 1996. Since 1996, he has been a Research Professor at the Centro de Investigaciones en Óptica, Leon, Mexico. He is National Researcher (SNI III), Mexico, and Regular Member of the Mexican Academy of Sciences. He has co-authored over 150 scientific papers and holds 4 patents. His research activity includes single-frequency, CW and Q-switched fiber lasers, laser dynamics, fiber sensors, and nonlinear fiber optics.

**Alexander V. Kir'yanov** received his Ph.D. degree in laser physics from the A.M. Prokhorov General Physics Institute (GPI), Russian Academy of Sciences (RAS), Moscow, Russia, in 1995. He has been with the GPI since 1987. Since 1998, he has been a Research Professor at the Centro de Investigaciones en Óptica (CIO), Leon, Mexico. He is National Researcher (SNI III), Regular Member of the Mexican Academy of Sciences, and Senior Member of the Optical Society of America. He has been co-author of over 200 scientific papers and holds 4 patents (Mexico, Russia, and USA); his current interests concern solid-state and fiber lasers and nonlinear optics of solid state and optical fiber.

**Jose L. Cruz** received his Ph.D. degree in physics from the University of Valencia, Valencia, Spain, in 1992. Initially, his career focused on microwave devices for radar applications; afterward, he joined the Optoelectronics Research Center, University of Southampton, Southampton, UK, where he was working in optical fiber fabrication. He is currently a Professor in the Department of Applied Physics, University of Valencia, where he is conducting research on fiber lasers and amplifiers, photonic crystal fibers, fiber gratings, microwave photonics and sensors.

**Miguel V. Andrés** is Professor at the Department of Applied Physics of the University of Valencia, Spain. He received his Ph.D. degree in physics from the University of Valencia, Spain, in 1985. Since 1983, he has successively served as Assistant Professor, Lecturer, and Professor in the Department of Applied Physics, University of Valencia, Valencia, Spain. After postdoctoral stay (1984-1987) at the Department of Physics, University of Surrey, UK, he has founded the Laboratory of Fiber Optics at the University of Valencia (www.uv.es/lfo). His current research interests include photonic crystal fibers, in-fiber acousto-optics, fiber lasers and new fiber-based light sources, fiber sensors, optical microcavities, microwave photonics, and waveguide theory. His research activity includes an increasing number of collaborations with Latin American universities and research institutes of Mexico, Argentina, and Brazil, among others. From 2006 to 2016, he was member of the External Evaluation Committee of the Centro de Investigaciones en Óptica, México. In 1999, he was awarded the Premio Cooperación Universidad-Sociedad 1999 of the Universidad de Valencia. Since 2009, he is a Member (Académico Correspondiente) of the Real Academia de Ciencias de Zaragoza.